\documentclass[twocolumn,showpacs,preprintnumbers,amsmath,amssymb]{revtex4}

\usepackage{graphicx}
\usepackage{dcolumn}
\usepackage{bm}

\begin{document}

\preprint{APS/123-QED}

\title{Dynamics of a microorganism moving by chemotaxis in its own secretion}

\author{Ankush Sengupta, Sven van Teeffelen, and Hartmut L\"owen}
\affiliation{%
Institut f\"ur Theoretische Physik II: Weiche Materie, 
Heinrich-Heine-Universit\"at \\
Universit\"atsstrasse 1, D-40225 D\"usseldorf, Germany 
}%

\date{\today}

\begin{abstract}

The Brownian dynamics of a single microorganism coupled by chemotaxis
to a diffusing concentration field which is secreted by the microorganism
itself is studied by computer simulations
in spatial dimensions $d=1,2,3$. Both cases of a chemoattractant and
a chemorepellent are discussed. For a chemoattractant, we find a transient
dynamical arrest until the microorganism diffuses for long times.
For a chemorepellent, there is a transient ballistic motion in all
dimensions and a long-time diffusion. These results are interpreted
with the help of a theoretical analysis.
\end{abstract}

\maketitle


\section{\label{sec:level1}Introduction}

Chemotaxis \cite{chemotaxis1,chemotaxis2,Jul} 
and Brownian motion \cite{doi86,dhont_book,pusey} belong to the 
key processes which govern the motility
of microorganisms (e.g. bacteria, amoeba
and endothelial cells) \cite{Poon_review}. 
In the simplest approach, the microbe ``smells" a chemical
and moves along the gradient of the concentration field of the chemoattractant
in order to reach efficiently the secretion source of the chemical. The opposite case 
of negative chemotaxis is realized in case the microbe intends to avoid 
another object which is secreting the chemical \cite{Bonner}. This chemotactic
drift is 
superimposed to stochastic motion due to fluctuations associated with
the active process of self-propulsion of the
microorganism \cite{geier-notes}. Chemotaxis can lead to clusters of
aggregated bacteria \cite{aggregated_clusters,geier2}
which are still emitting chemoattractant.

Here we study the self-coupled situation where the microorganism ``smells" 
itself, i.e. it reacts chemotactically to its own secreted chemical. This ``autochemotaxis"
is realized in aggregated clusters of different bacteria if the aggregate is considered 
as a net particle. Another realization is a single bacterium which has both an emitter and a sensor of the
same chemical. Tsori and de Gennes \cite{deGennes_2004} have studied a simple model for
this situation in different spatial dimensions $d$ and find self-trapping of the bacterium
 in its own chemoattractant cloud for $d=1,2$ but not for $d=3$. 
This means that for low dimensionality 
 the bacterium is fooled by its own secretion such that it is getting
localized for long times.
In a subsequent numerical study
of a model microbe coupled to its own chemoattractant secreted at constant rate, 
Grima \cite{Grima_PRL_2005,Grima_PRE_2006}
calculated the long-time dynamics in various dimensions $d$ and found long-time diffusive behavior
even for $d=1,2$ in contradiction to Ref.\ \cite{deGennes_2004}. Grima also studied the case of negative chemotaxis
and finds in all dimensions long-time diffusion or ballistic motion depending on the strength $\lambda<0$
which couples the microbe's driving force to the gradient of the chemical concentration field.
Grima predicts that if  $|\lambda|$ exceeds a critical value $\lambda_c$, then the long-time motion is super-diffusive.

In this paper we revisit autochemotaxis by studying a model which is similar 
but not identical to that proposed by Grima
 \cite{Grima_PRL_2005}. In the model of Grima, a finite global extinction rate
($\Lambda$) of the secreted chemical is present. Here we focus on the
case $\Lambda=0$, which can be realized in experiments (see e.g. \cite{Bonner}).
By extensive computer simulations, we study
 different spatial dimensions $d=1,2,3$ and both cases of positive and negative chemotaxis. 
Our results are summarized as follows:
Consistent with Grima \cite{Grima_PRL_2005}, we find for positive chemotaxis
(i.e.\ for a coupling parameter $\lambda>0$) long-time diffusive motion. In addition we find 
{\it dynamical transients\/} before reaching the long-time diffusive limit. During the transients, 
the dynamics of the microorganism is strongly reduced resulting in almost dynamical arrest.
The crossover time from intermediate arrest to long-time diffusion grows strongly  with  the coupling $\lambda$. Therefore the idealized analysis 
of Tsori and de Gennes who predicted self-trapping (i.e.\ a complete dynamical arrest)
is manifested by long transients for very strong couplings \cite{selftrapping}.
The averaged mean-square displacement
as a function of time does not exhibit a universal slope in this transient regime, but the 
actual mean slope is decreasing with an increase in the coupling $\lambda$. The transient behavior is most pronounced
in one dimension but weakened considerably in three dimensions. 

For negative chemotaxis (i.e. dynamical self-avoiding of the microbe),
on the other hand,
we find transient ballistic motion in all dimensions. In $d = 2,3$ we
observe a long-time diffusion for all coupling strength.
Such long-time diffusive motion in $d=1$ , though not directly
observed in the simulation, is possible by finite probability of
changing the direction of motion (left to right) at long time scales
for non-zero temperature.
According to Refs.\ \cite{Grima_PRL_2005}, the critical 
coupling $\lambda_c$ above which ballistic long-time behavior is found
depends on the global extinction rate $\Lambda$ but stays finite when $\Lambda \to 0$.
One reason for the discrepancy is because noise has not been completely
included in the earlier treatments \cite{Grima_PRL_2005},
while solving for the integrals concerning the non-Markovian chemotactic
force. We take note of the effect of noise on the
microbe's trajectory in an appropriate place in this paper (for another
example demonstrating the importance of noise, see Ref.~\cite{geier-notes}).
Again we address the transient behavior
and find an intermediate time window where the super-diffusive
motion is found between a short-time and 
long-time diffusive behavior. This motion we observe is similar to persistent
random walk of a microbe which can be mapped to the worm-like chain model
\cite{Poon_review,Friedrich08}. In the particular case we consider,
the persistence length is seen to depend on the microbe's coupling with the
repellent.

Our predictions can in principle be verified in experiments on aggregates
of bacteria. 
For many bacteria our model reduces to particles interacting via
gravitation-kind potentials for $d=3$. Therefore our analysis might have
applications for Brownian dynamics of gravitational
matter \cite{Acedo_2006,ChavanisPRE08,ChavniasPRE05}.
Further generalizations of our model are to predator and prey models
possibly leading to interesting spatiotemporal delay effects, see e.g.\ \cite{Trimper}.

Our paper is organized as follows: in section II, we propose the model
of a microorganism coupled
to its own chemoattractant/chemorepellent, provide the simulation details
and point out experimental situations to compare typical estimates of
parameters used in the model.
In section III we present the results of our investigation. In section IV
we explain our findings with simple theoretical
analysis. We conclude the paper in section V by discussing the main points
of our findings, comparisons of our results with relevant practical cases
of self-propulsion, and future directions of our research.


\section{\label{sec:level2}The Model and Simulation Details}

{\em Model:} The microorganism is
modeled as a point particle which undergoes completely overdamped
Brownian motion with an effective temperature $\beta^{-1}$
in a medium with viscosity coefficient $\gamma$. This `particle'
is assumed to emit a chemical, with which it is self-coupled,
continuously in time.
In this context, the word `particle' is thereby taken to represent
the chemotactic agent of interest - an idealized microorganism coupled with a
self-emitted chemical field. Fluctuations associated with the
active process of self-propulsion of the microbe is modeled
by the effective temperature parameter $\beta^{-1}$ in our system.
The time evolution of the density field $\rho({\bf{r}},t)$ of the
continuously emitted chemical is thus governed by a diffusion equation
with a source term which depends upon the instantaneous position
${\bf r}_{b}({t})$ of the moving microbe:
\begin{equation}
\frac{\partial \rho({\bf{r}},t)}{\partial t} = 
D_{c}{\nabla}^2 \rho({\bf{r}},t) 
+ \lambda_{e} \delta({\bf r}-{\bf r}_{b}({t})) .
\label{eq:rd}
\end{equation}
Here, the constants $\lambda_{e}$ and $D_c$ are the rate of emission
of the chemical and the diffusion constant of the chemical in the medium,
respectively.

In the absence of chemical, the microbe diffuses non-chemotactically in
the medium with an effective free diffusion constant
$D = 1/(\gamma \beta)$. However, with the presence of the
emitted chemical the resulting `chemotactic' behavior
depends on the nature of the self-coupling of the microorganism
to its chemical field, i.e. whether it moves `up' or `down' the
chemical density gradient.
We study both
cases by simply modeling the self-coupling `force' to be proportional
to the gradient of the chemical density field $\nabla \rho ({\bf r},t)$,
and the proportionality
constant $\lambda$ determines the strength as well as the nature of
the coupling.
In reality, of course, chemotaxis can be more complex involving
temporally sampling of the concentration field and a 
biased random walk \cite{grant,kafri,Poon_review}. In this context, we
note that the `chemotactic force' imitates the net effect of chemotactic
movements on a phenomenological level. The details of the actual propulsion
mechanism \cite{shapere} of the chemotactic agent and the
effect of the solvent flow field on the diffusing chemical are not taken
into account. Within our simple model,
positive and negative $\lambda$ naturally generate the cases of
positive and negative chemotaxes, respectively. 
We are thus lead to the following idealized model of a chemotactic agent
\begin{equation}
\gamma{\bf {\dot r}}_{b} (t) = {\bf F}({\bf r}_{b},t)
+ {\mbox {\boldmath $\eta$}}(t) .
\label{eq:bd}
\end{equation}
Here, {\boldmath ${\eta}$}$(t)$ is an effective noise
specified by $\langle{\mbox {\boldmath $\eta$}}(t)\rangle = {\bf 0}$ and
$\langle{\mathbf \eta_{i}}(t)
{\mathbf \eta_{j}}(t^{\prime}) \rangle = 
2\gamma\beta^{-1}\delta_{ij}\delta(t-t^{\prime})$, with $i$ and
$j$ referring to the spatial components of the noise vector.
This noise term is assumed to effectively take care of all non-equilibrium
fluctuations that may be associated with the {\em active} process
\cite{TonerTuRamaswamy,geier1} of self-propulsion, in absence
of the chemical.
${\bf F}({\bf r}_{b},t)$ denotes the model `chemotactic force' taken to
imitate the systematics of the effective chemotactic movement of the
microbe at the position ${\bf r}_b$ at time $t$ due to the chemical
secreted all along
the trajectory traversed in the past. It is obtained by analytically
solving Eq.~(\ref{eq:rd}) for the density field $\rho ({\bf r},t)$
by the method of Green's function, and subsequently calculating the
gradient $\nabla \rho ({\bf r},t)$.
The `force' at a time instant is dependent on the entire previous path
history of the microorganism, thereby generating a strongly non-Markovian
dynamics. However, owing to a physical {\em memory time}
($t_0$) associated with the microbe to sense its chemical,
the part of the trajectory in this most recent time $t_0$, i.e. for
all ${\bf r}_b (t^{\prime})$ with $t-t_0 < t^{\prime} \le t$,
does not contribute. The physical import
of this is that there is a finite time delay $t_0$, however small,
between the act of secreting chemical by the microorganism and the act of
responding to it, during which the sensor gets to activate.
With the introduction of
the memory time, $t_0$, the `chemotactic force' at time $t$ and at position
${\bf r}$ becomes
\begin{equation}
{\bf F}({\bf r},t) = -2 \lambda \lambda_{e} \int_{0}^{t-t_0}dt^{\prime}
{\frac{({\bf r}-{\bf r}_{b}(t^{\prime}))}{4D_{c}|t-t^{\prime}|}}
\frac{\exp[\frac{-({\bf r}-{\bf r}_{b}(t^{\prime}))^2}{
(4 D_{c}|t-t^{\prime}|)}]}{(4\pi D_{c}|t-t^{\prime}|)^{d/2}} ,
\label{eq:gradrho1}
\end{equation}
where $d$ is the dimensionality of the embedding space. Evidently
form Eq.~(\ref{eq:gradrho1}), for the mathematical case of $t_0 = 0$,
the `force' becomes divergent.

{\em Simulation details:}
We performed extensive Brownian dynamics simulation
\cite{allen_tildesley_book} for this
non-Markovian process of a microorganism moving by autochemotaxis.
We measured time in units of $\tau_0 = \lambda_{e}^{-1}$, all lengths in units
of $l_{0} = ({\sqrt{D D_c}}/\lambda_e )^{1/2}$ and energies in units
of $\beta^{-1}$.
The coupling strength $\lambda$ is measured in units of $\beta^{-1} {l_0}^d$.
Thus, we set $\lambda_{e}=1$, $l_0 = 1$ and $\beta = 1$ for convenience.
Further we
considered the physical situation when the microorganism diffuses at
a much slower
rate compared to the emitted chemical in the medium \cite{Hofer,Luca},
taking $D=0.1 \ {l_{0}}^{2} /\tau_{0}$ and fixing the ratio $D/D_{c} = 0.01$.
In our Brownian dynamics simulations we used $t_0 = 0.001 \ \tau_{0}$.
The Langevin equation [Eq.~(\ref{eq:bd})] is solved with a discrete
time step $\Delta t = 0.0001 \ \tau_{0}$. Space is, however, continuous.

{\em Connection to experiments:}
In order to get an estimate for the coupling
strength $\lambda$ in our units, we note that the typical value
of the ejection rate of chemical from a microorganism is
$\lambda_e \sim 10^3 \ {\rm molecules} / s$ \cite{deGennes_2004,Brenner},
and usually $D/D_c \sim 10^{-1} - 10^{-2}$ \cite{Hofer,Luca}.
In three spatial dimensions, for example,
the chemotaxis of {\em Dictyostelium} to shallow cAMP gradients
\cite{Endres,HaastertPostma} with typical values of
$\nabla \rho \sim 0.01 \ nM / \mu m$,
$D_c \sim 300 \ \mu m^2 /s$, $D/D_c \sim 10^{-2}$, and
moving with steady state velocity $v \sim 0.2 \ \mu m/s$,
yields $\lambda \sim 10^{4} \ \beta^{-1} {l_0}^3$. 

Chemotactic {\em Microglia} cells \cite{Luca}
move at a speed of $v \sim 2 \mu m /{\rm min}$ in a spatial gradient
$\nabla \rho \sim 0.003 \ nM / \mu m$ of a chemoattractant
(${\rm {IL-1\beta}}$) which is secreted at a rate of
$\lambda_e \sim 200 \ {\rm molecules} / {\rm min}$ and diffuses with
$D_c = 900 \ \mu m^2 /{\rm min}$. The coupling strength in this case
is $\lambda \sim 10 \ \beta^{-1} {l_0}^3$ for an effective non-chemotactic
diffusion constant $D \sim 33 \ \mu m^2 /{\rm min}$ of microglial cells,
due to random motility in the tissue. The chemoattractant has a low
decay rate of $\Lambda \sim 0.003 - 0.03 \ {\rm min}^{-1}$.
Microglial cells are also known to respond to a chemorepellent
(${\rm {TNF-\alpha}}$) they produce, with similar production,
diffusion and decay rates as the chemoattractant. Our estimate of
the chemoattractant gradient was based on the value of the
chemotactic coefficient
$\sim 780 \ \mu m^2 \ nM^{-1} \ {\rm min}^{-1}$, a ratio between cell
velocity and chemical gradient, used in Ref. {\cite{Luca}}.
The corresponding value for the repellent is not known.

Again, for {\em E. coli} of size $\sim 1 \ \mu m$, swimming
at $v \sim 20 \ \mu m/s$ in the
spatial gradient $\nabla \rho \sim 0.1 \ \mu M / \mu m$ of a chemoattractant
diffusing with $D_c = 10^{-5} \ cm^2 /s$,
in a medium of viscosity $10^{-3} \ {\rm Pa}.s$ \cite{Schnitzer,BergPurcell};
the coupling strength is $\lambda \sim 10^{-1} \ \beta^{-1} {l_0}^3$. The
non-chemotactic diffusion coefficient of the bacterium is
$D = 6.6 \times 10^{-6} \ cm^{2} /s$. The time required by the chemical
in this case, to diffuse a length equal to the size of the bacterium,
is of the order of $0.1 \ \tau_0$. The memory time, $t_0$, needed by the
bacterium to respond to the chemical stimulus can be much smaller than
this time.

In all the above calculations, the effective non-chemotactic
diffusivity $D$ was used to express the energy unit $\beta^{-1} = \gamma D$.

\begin{figure*}[]
\includegraphics[width=16.0cm]{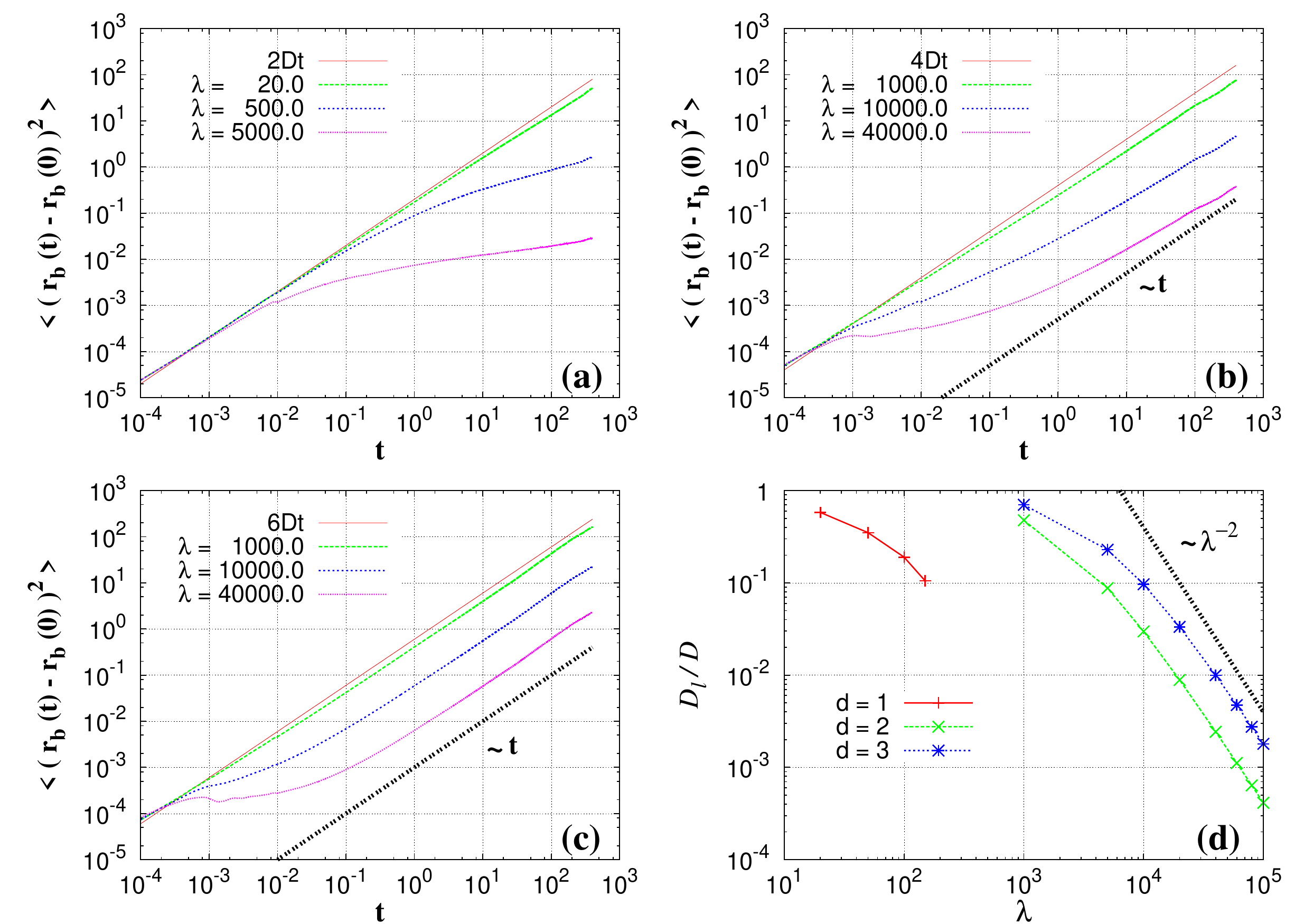}
\caption{\label{figAttrR2} (color online). Mean square displacement
$\langle ({\bf r}_b (t) - {\bf r}_b (0))^2 \rangle$
of the microorganism as a function of time $t$ with chemoattractant
in (a) $d = 1$ with $\lambda = 20, 500, 5000$;
(b) $d = 2$ with $\lambda = 1000, 10000, 40000$;
(c) $d = 3$ with $\lambda = 1000, 10000, 40000$.
The non-chemotactic diffusion reference lines are also indicated as
$2Dt$, $4Dt$ and $6Dt$ correspondingly for $d = 1, 2, 3$. Reference
lines (thick dotted) are used to indicate the long-time diffusive
behavior ($\sim t$) wherever possible. The relative
long-time diffusivity $D_{l}/D$ is shown as a
function of $\lambda$ in (d) for $d = 1,2, 3$.The reference line (thick
dotted) shows a power law scaling behavior $\sim 1/\lambda^2$ (see text).
}
\end{figure*}

\begin{figure*}[]
\includegraphics[width=16.0cm]{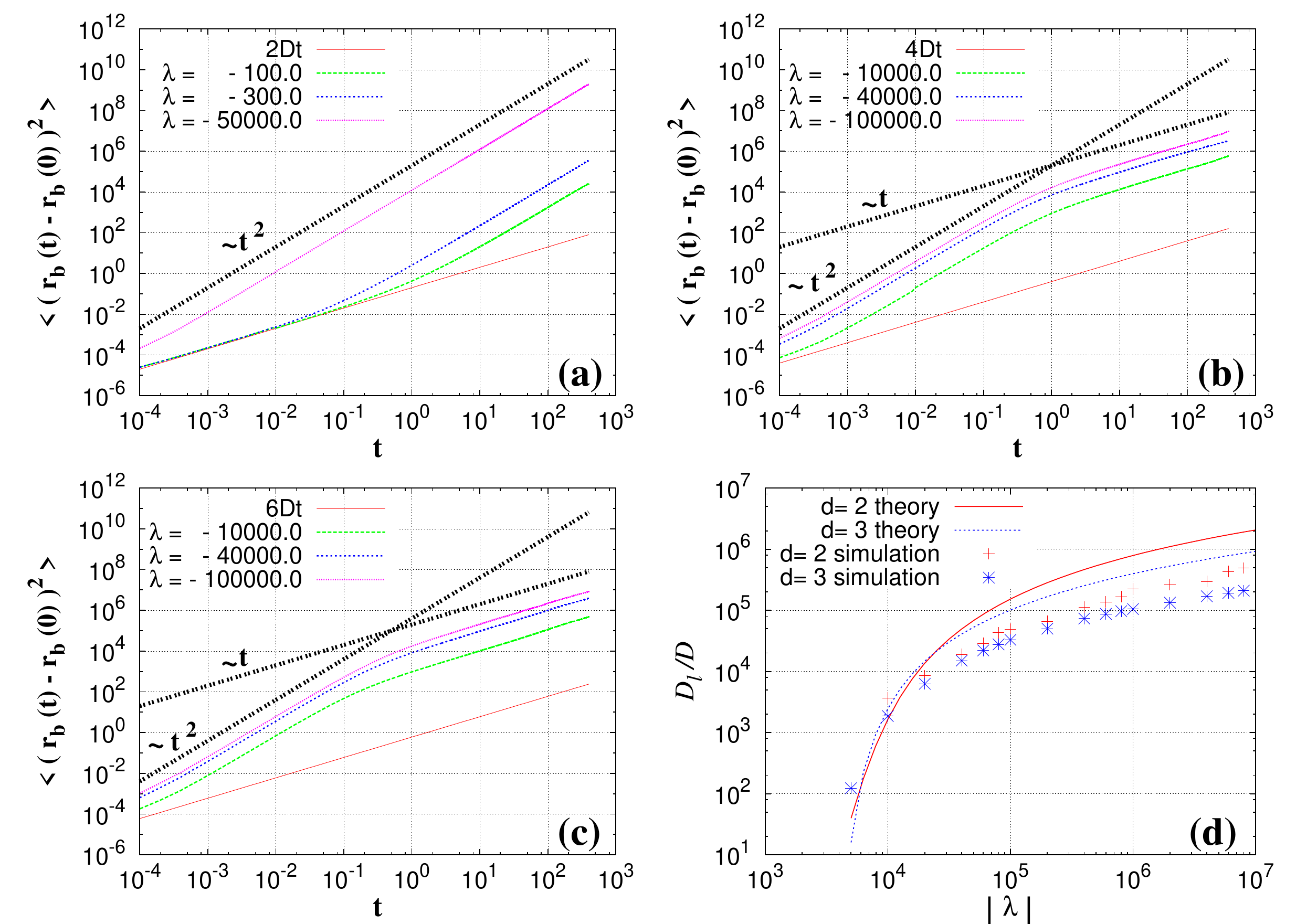}
\caption{\label{figReplR2} (color online). Mean square displacement
$\langle ({\bf r}_b (t) - {\bf r}_b (0))^2 \rangle$
of the microorganism as a function of time $t$ with chemorepellent
in (a) $d = 1$ with $\lambda = -100, -300, -50000$;
(b) $d = 2$ with $\lambda = -1000, -40000, -100000$;
(c) $d = 3$ with $\lambda = -1000, -40000, -100000$.
The non-chemotactic diffusion reference lines are also indicated as
$2Dt$, $4Dt$ and $6Dt$ correspondingly for $d = 1, 2, 3$. Reference
lines (thick dotted) indicating the ballistic ($\sim t^2$) and the
long-time diffusive ($\sim t$) dynamics are shown as guide to the eye.
The relative long-time diffusivity $D_{l}/D$ is shown as a
function of $|\lambda|$ in (d) for $d = 2, 3$. The points represents
the actual data obtained from simulations, the lines correspond
to a semiquantitative theory (see text).
}
\end{figure*}

\section{\label{sec:level3}Results}
 
We now investigate the nature of the dynamics
in all the dimensions and for both cases of positive and negative
$\lambda$, examining the model microorganism from some initial reference point
taken as the origin, i.e. ${\bf r}_b (t=0) = 0$. For this purpose we
computed the mean square
displacement of the microbe as a function of time and averaged over
$10^3$ realizations for each case. We checked that the system is in
a steady state, and have also performed a steady state averaging of
the mean square displacement. We illustrate our findings below.

\subsection{\label{sec:level3}Positive autochemotaxis}


Upon examining the motion of the microbe in a chemoattractant ($\lambda > 0$)
in one dimension ($d=1$), we found signatures of long-time diffusion
($\langle ({\bf r}_b (t) - {\bf r}_b (0))^2 \rangle \sim t$) with a
modified diffusion constant $D_{l}$. The value of $D_{l}$ depends on
the strength of the coupling $\lambda$, and decreases with increasing
$\lambda$. For very high $\lambda$ values it is computationally difficult
to obtain the long time diffusive behavior; but we obtained an upper
estimate of the diffusivity for lower $\lambda$ values from a fit to
the obtained data.
In the opposite limit, i.e. at very short times, the microbe's dynamics
is also diffusive with the non-chemotactic diffusion constant $D$. The departure
from this behavior occurs at times dependent on the coupling strength:
the stronger the coupling, the dynamics become markedly history dependent
and the microorganism deviates from the non-chemotactic diffusion faster.
At intermediate times we find long period of sub-diffusive crossover
regime, showing signatures of a
transient dynamical arrest at high $\lambda$-values. The crossover
time also increases with the coupling strength. Fig.~\ref{figAttrR2}(a)
shows the mean square displacement as a function of time in $d=1$,
for three values of $\lambda$ and compared with the non-chemotactic
diffusion.

In $d=2$, we find similar long-time diffusive behavior. The
crossover times from early-time non-chemotactic diffusion to the
long-time modified diffusion, is greatly reduced for a given $\lambda$-value
as compared to the one dimensional case. This feature is attributed
to the effect of fluctuations. The crossover time to the final long-time
diffusion, however, increases
with coupling strength. The intermediate sub-diffusive regime is
also diminished in time. The mean square displacement as a function
of time, in this case, is shown in Fig.~\ref{figAttrR2}(b).

In Fig.~\ref{figAttrR2}(c) we show the case for $d=3$, and in consistence with
our expectation we find similar long-time diffusion with a further reduced
crossover time as compared to the low-dimensional cases. Thus the
transient dynamical arrest becomes more prominent at lower dimensions
and at stronger couplings.
The $\lambda$-dependence of the modified diffusion constant
$D_{l}$, is shown in Fig.~\ref{figAttrR2}(d), for all dimensions,
relative to the corresponding non-chemotactic diffusion constant $D$.

\subsection{\label{sec:level3}Negative autochemotaxis}

For the repulsive case, when the model microorganism gets repelled by its
ejected chemical ($\lambda < 0$), we found a short-time non-chemotactic
diffusive motion with a crossover to a ballistic behavior
($\langle ({\bf r}_b (t) - {\bf r}_b (0))^2 \rangle \sim t^2$) in one
dimension, for all values of the coupling strength
(Fig.~\ref{figReplR2}(a)). The time of commencement of
the ballistic behavior, however, depends on $\lambda$. For weak coupling
(i.e. low $|\lambda|$), the dynamics resembles the non-chemotactic
diffusion for longer times before finally going over to the ballistic
dynamics. The velocity of this ballistic motion is given by the
time derivative of the root mean square displacement,
${\frac{\partial}{\partial t}}\sqrt{\langle ({\bf r}_b (t) - {\bf r}_b (0))^2 \rangle}$,
and this velocity is seen to increase with increase of the coupling
strength $|\lambda|$.
We argue that the microorganism can change the direction of its motion
in one dimension due to a non-vanishing finite barrier crossing probability.
Under such circumstance the motion will be diffusive
at very long times (not seen in the simulation), with higher diffusion
constant.

In higher dimensions, $d=2$ (Fig.~\ref{figReplR2}(b)) and
$d=3$ (Fig.~\ref{figReplR2}(c)), we observed a crossover from the
non-chemotactic diffusion to a long-time diffusion for all
coupling strengths, with
a ballistic transient dynamics at intermediate times for
high $|\lambda|$. The velocity of the transient ballistic motion
increase with increasing coupling strength.
The time duration of this ballistic transient as well
as the modified long-time diffusion constant are also $\lambda$-dependent,
both increase with increase in $|\lambda|$. In {Fig.~\ref{figReplR2}(d)} we
show this modified diffusion constant for the
repulsive case relative to the non-chemotactic diffusion constant as
a function of $|\lambda|$, in $d=2, 3$.


\section{\label{sec:level3}Theory}

Our findings can be understood qualitatively and partly quantitatively
by relatively simple theoretical considerations, which are presented
in the current section.  In subsection~\ref{subsec:theory_pos}, a
scaling law for the diffusion constant for the case of positive
chemotaxis, $D_{l}\propto D_c^{d+2}t_0^{d-2}/\lambda^2$, is derived
based on a simple rate theory. The same scaling
$D_{l}\propto\lambda^{-2}$ had been predicted for the slightly
different model , which includes evaporation, by Newman and
Grima~\cite{NewmanGrima_PRE_04} and Grima~\cite{Grima_PRL_2005}. In
subsection~\ref{subsec:theory_neg} we present a theory, which
quantitatively predicts the long-time diffusion constant in the case
of negative chemotaxis for all coupling strengths.

Both approaches, for negative and positive chemotaxis, attribute the long-time
diffusion to small fluctuations about the respective steady states in the case
of zero noise. In each subsection we therefore first present the solutions to
the equation of motion~(\ref{eq:bd}) without fluctuations (i.e., at zero
temperature or infinite coupling strength $\lambda$), before discussing the
influence of small fluctuations on the long-time dynamics.

\subsection{Positive autochemotaxis}
\label{subsec:theory_pos}%
In the case of strong positive autochemotaxis the model microorganism
is trapped within
its own secretion, which effectively provides an attractive external potential
at the microbe's location ${\bf r}_b(t)$. For zero noise, the microbe is at
rest, i.e., ${\bf r}_b(t)={\bf r}_b$, and does not experience any force. If,
on the contrary, the microbe was at time $t$ instantaneously placed a
distance ${\bf r}-{\bf r}_b$ away from the location, which it occupied at all
earlier times $t'<t$, it would feel a force ${\bf F}_s({\bf r})$, which is
obtained analytically by evaluating Eq.~\eqref{eq:gradrho1} (see also
Fig.~\ref{fig:forceattr}):
\begin{figure}
\includegraphics[width=8.0cm]{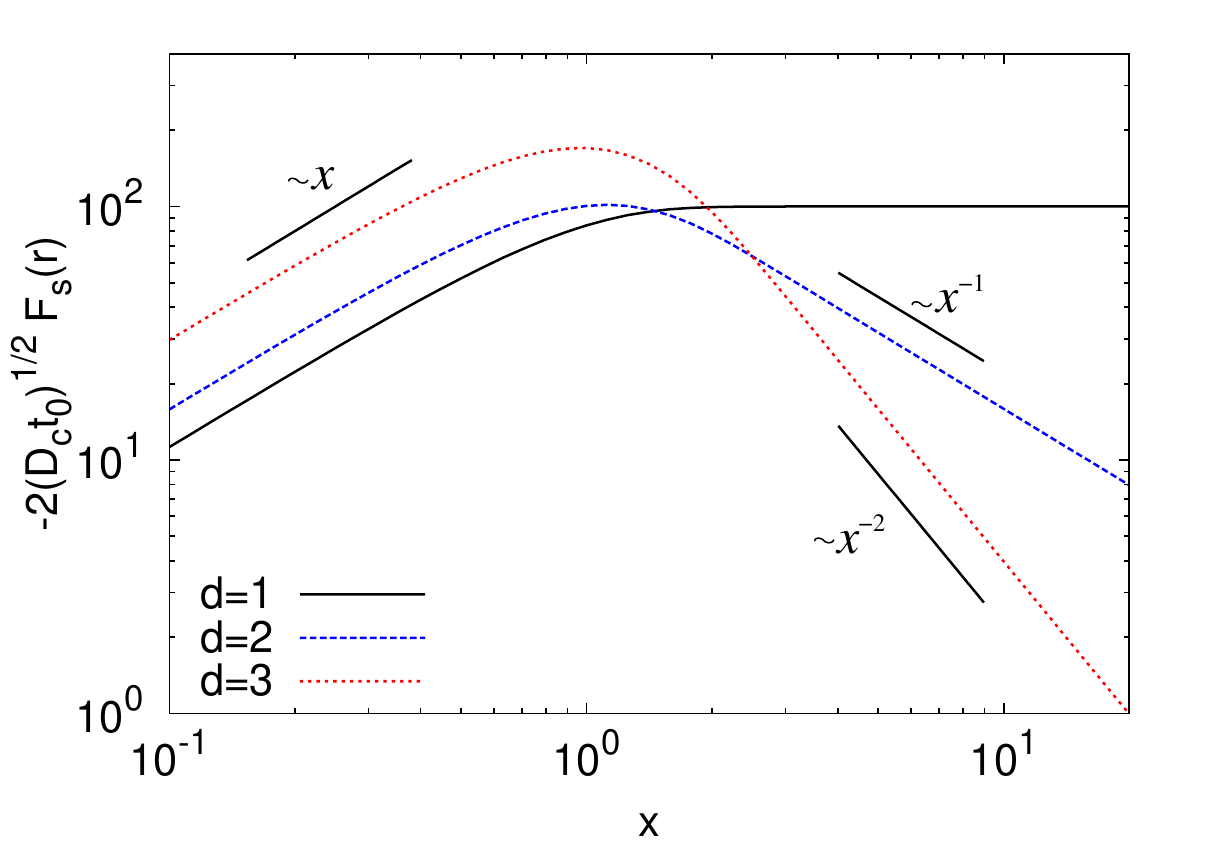}
\caption{\label{fig:forceattr} (color online). The force $|{\bf F}_s({\bf r})|$, which a
  microorganism would experience if it was at time $t$ instantaneously placed a
  distance ${\bf r}-{\bf r}_b$ away from the location, which it occupied at
  all earlier times $t'<t$, plotted as a function of $x=|{\bf r}-{\bf
    r}_b|/(2\sqrt{D_c t_0})$ for all dimensions $d=1,2,3$. Representative
  parameters chosen are $D_c=0.1$, $t_0=0.01$, and $\lambda=10^4$.}
\end{figure}
\begin{equation}\label{eq:fss}
{\bf F}_s({\bf x})=\left\{\begin{array}{ll}
-\frac{\lambda}{2D_c} {\rm erf}\left(x\right)
\frac{{\bf x}}{x}\,,&d=1\\
-\frac{\lambda}{4\pi D_c^2t_0} \left(1-e^{-x^2}\right)
\frac{{\bf x}}{x^2}\,,&d=2\\
-\frac{\lambda}{4\pi^{3/2} D_c^2t_0} 
\left[\frac{\sqrt{\pi}}{2x}{\rm erf}\left(x\right)
-e^{-x^2}\right]\frac{{\bf x}}{x^2}
\,,&d=3\,.
\end{array}\right.
\end{equation}
Here, ${\bf x}=({\bf r}-{\bf r}_b)/(2\sqrt{D_c t_0})$ is the dimensionless
distance from the former position ${\bf r}_b$ of the microbe and $x=|{\bf x}|$
is its absolute value.  Locally, i.e., for distances $|{\bf r}-{\bf r}_b|\ll
2\sqrt{D_c t_0}$, the force is linear in the distance and given by
\begin{equation}\label{eq:fsslimit}
|{\bf F}_s({\bf r})|\simeq\left\{\begin{array}{ll}
\left(2\sqrt{\pi D_c^3 t_0}\right)^{-1}\lambda |({\bf r}-{\bf r}_b)|\,,&d=1\\
\left(8 \pi D_c^2 t_0\right)^{-1}\lambda |({\bf r}-{\bf r}_b)|\,,&d=2\\
\left(24 \sqrt{\pi ^3 D_c^5 t_0^{3}}\right)^{-1}\lambda |({\bf r}-{\bf r}_b)|\,,&d=3\,.
\end{array}\right.
\end{equation}

Without fluctuations, the microbe never experiences the described force
field. However, if the noise is nonzero but small, the microbe eventually
walks up the locally parabolic walls of its self-generated potential for a
short time $\tau$, which, in turn, might lead to a shift of the position of
the minimum ${\bf r}_0$ of the self-generated attractive potential by a
distance $\Delta{\bf r}_0={\bf r}_0(t+\tau)-{\bf r}_0(t)$. Clearly, the
minimum is only displaced, if the duration of the excursion $\tau$ is larger
than the memory time $t_0$ and if the microbe excurses predominantly in one
direction. After time $t+\tau$ the microbe position might relax to the new
minimum, ${\bf r}_b(t'>t+\tau)\rightarrow{\bf r}_0(t+\tau)$. By such a
fluctuation, the microorganism effectively manages to move a
distance $\Delta{\bf r}_0$ within the time $\tau$.

Most relevant for the long-time diffusion are those fluctuations, which yield
a large displacement $\Delta{\bf r}_0$ and still occur at a high rate
$\gamma_{R} \lesssim 1/\tau$. These fluctuations are regarded to constitute
the relevant mean steps in an effective continuous time random walk with the
desired diffusion constant $D_{l}\sim\Delta{\bf r}_0^2/\tau$.

In this subsection, we only attempt to obtain a scaling law for the
long-time diffusivity. We therefore restrict our consideration to a
subset of very simple fluctuations, which are believed to be
representative for all fluctuations relevant for diffusion. In
particular, we consider a ``jump'' process at time $t=0$: At all
earlier times, $t<0$, the microbe is resting at the location of the
self-generated potential's minimum at the origin, i.e., ${\bf r}_b(t
<0)={\bf r}_0(0)={\bf 0}$, before it undergoes an excursion to a new
location ${\bf r}_b(\epsilon<t<\tau)=\Delta{\bf r}_b$ within a time
$\epsilon$; the latter timescale is assumed to be small with respect
to the residence time $\tau$. During the latter time span $\tau$ the
microbe stays at the new position, where it resists the force due to
the secretion from earlier times.  In the case of a large coupling
strength $\lambda$ the process of getting to the new location during
the time span of duration $\epsilon$ is irrelevant. After time $\tau$
the position of the model microbe is deterministically relaxing towards
the new minimum of the chemoattractant at $\Delta{\bf r}_0(\tau,\Delta{\bf
  r}_b)<\Delta{\bf r}_b$, which is a function of the parameters $\tau$
and $\Delta{\bf r}_b$, only.  Summarizing, the simple pathway is
described by

\begin{equation}\label{eq:simpleprocess}
{\bf r}_b(t)=\left\{\begin{array}{ll}
{\bf r}_b(0)={\bf 0}\,,&t<0\\
\Delta{\bf r}_b\,,&\epsilon<t<\tau\\
\Delta{\bf r}_0(\tau)\,,&t>\tau\,.
\end{array}\right.
\end{equation}

The probability of the described fluctuation is proportional to the
Arrhenius factor \cite{risken}
\begin{equation}\label{eq:arrhenius}
  p\left[\tau,\Delta r_b\right]
  \propto \exp\left[-S(\tau,\Delta r_b)/4\right]\,,
\end{equation}
with $\Delta r_b=|\Delta{\bf r}_b|$ and $S(\tau,\Delta r_b)$ the
Onsager-Machlup action
\begin{equation}\label{eq:om}
S(\tau,\Delta r_b)=\gamma\int_0^{\tau}{\rm d}t\left|\dot{\bf r}_b(t)-\gamma^{-1}{\bf F}(t)\right|^2\,.
\end{equation}
Neglecting the initial process of moving the distance $\Delta r_b$ and also
ignoring the relaxation of the chemoattractant during the time $\tau$ the
action is approximately given by

\begin{equation}\label{eq:om2}
S(\tau,\Delta r_b)\approx\gamma^{-1}\tau\left[{\bf F}_s(\Delta {\bf
    r}_b)\right]^2\,,
\end{equation}
which, together with Eqs.~(\ref{eq:fss}) and (\ref{eq:fsslimit}) yields
\begin{equation}\label{eq:om2approx}
S(\tau,\Delta r_b)\propto\left\{\begin{array}{ll}
\frac{\lambda^2\Delta r_b^2\tau}{\gamma D_c^2(D_ct_0)^d}\,,&\Delta r_b\ll 2\sqrt{D_ct_0}\\
\frac{\lambda^2\tau}{\gamma D_c^2\Delta r_b^{2(d-1)}}\,,&\Delta r_b\gg 2\sqrt{D_ct_0}\,.
\end{array}\right.
\end{equation}
%

According to Eq.~\eqref{eq:arrhenius}, a minimum requirement for the
described fluctuation to occur frequently is that $S(\tau,\Delta r_b)$
does not exceed a value of the order of $1$, i.e., $S(\tau,\Delta
r_b)\lesssim1$. As Eq.~(\ref{eq:om2approx}) is a strictly
monotonically increasing function of $\tau$ and $\Delta r_b$, this
constraint is equivalent to the equality
\begin{equation}
  \label{eq:constraint_action}
  S(\tau,\Delta r_b)=1\,.
\end{equation}
Diffusion is believed to be governed by those random displacements of
the microorganism, which maximize the shift of the potential's minimum
$\Delta r_0(\tau,\Delta r_b)=|\Delta{\bf r}_0(\tau,\Delta {\bf r}_b)|$
subject to this constraint.

We will shortly see that for large coupling constants $\lambda^2\gg
D_c^{d+1}t_0^{d-2}\gamma$ small excursions $\Delta r_b\ll
2\sqrt{D_ct_0}$ are most relevant. In this latter limit, $\Delta r_0$
is obtained analytically as
\begin{equation}
  \label{eq:potentialminimum_smalldeltarb}
  \Delta r_0(\tau,\Delta r_b)\simeq \left[1-\left(t_0/\tau\right)^{d/2}\right]\Delta r_b\,.
\end{equation}
The maximum of $\Delta r_0$ subject to the constraint of
Eq.~(\ref{eq:constraint_action}) is given by 
\begin{equation}  \label{eq:optimum_shift}
\Delta r_0^*\simeq\frac{d}{d+1}\Delta r_b^*\,,
\end{equation}
at an optimum displacement of the microbe and an optimum residence time
\begin{eqnarray}\label{eq:optimum_displacement_restime}
\Delta r_b^*&\simeq&a\sqrt{\gamma D_c^{d+2}t_0^{d-1}}/\lambda\,,\nonumber\\
\tau^*&\simeq&b t_0\,,
\end{eqnarray}
with prefactors $a=2^{-1}$, $3^{-1/2}$, $2^{-2/3}$ and $b=4$, $3$,
$2^{4/3}$, in one, two, and three dimensions, respectively. Clearly,
Eq.~(\ref{eq:optimum_displacement_restime}) fulfills the before
mentioned assumption of small excursions in the limit of large
$\lambda^2\gg D_c^{d+1}t_0^{d-2}\gamma$.
Eq.~(\ref{eq:optimum_displacement_restime}) yields the desired scaling
behavior for the long-time diffusivity $D_{l}\propto \Delta
r_0^2/\tau$,
\begin{equation}\label{eq:longtimediffusion_attr}
D_{l}\propto\frac{\gamma D_c^{d+2}t_0^{d-2}}{\lambda^2}\,.
\end{equation}
In conclusion, the simple theory predicts an inverse quadratic
dependence of $D_{l}$ on $\lambda$ in the strong-coupling limit, which
has also been observed in the computer simulations [see
Fig.~\ref{figAttrR2}(d)]. The same scaling had already been predicted
for the slightly different model, which includes evaporation, by
Newman and Grima~\cite{NewmanGrima_PRE_04} and
Grima~\cite{Grima_PRL_2005}.

\subsection{Negative autochemotaxis}
\label{subsec:theory_neg}
\begin{figure}
\includegraphics[width=8.0cm]{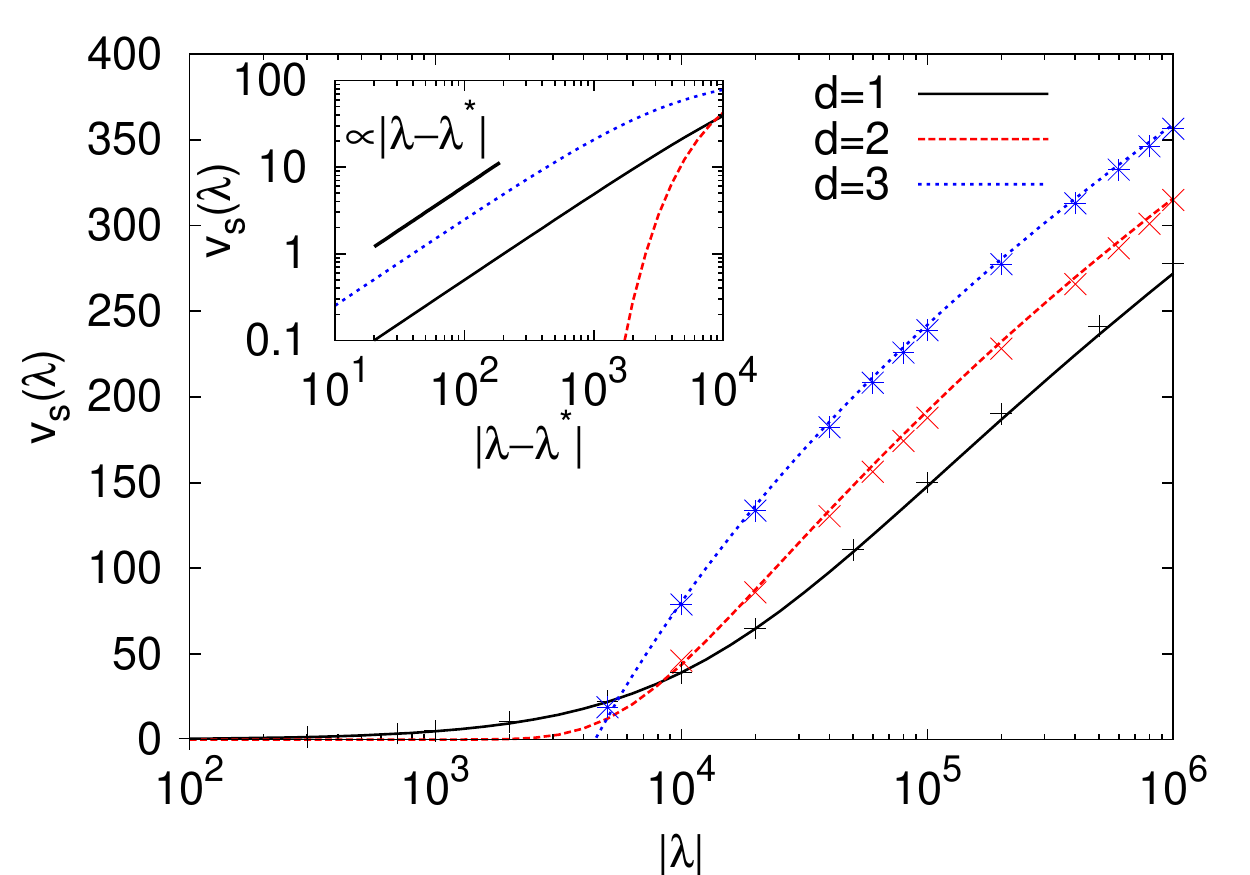}
\caption{\label{fig:steadystate} (color online). The steady-state velocity
  $v_s(\lambda)$ as a function of coupling constant $|\lambda|$ for one,
  two, and three spatial dimensions, determined theoretically (lines),
  compared to the square root of the slopes of the mean-square
  displacements at early times,
  ${\partial{\sqrt{\langle({\bf r}_b(t)-{\bf r}_b(0))^2}}\rangle/\partial t}$,
  for two and three dimensions
  (symbols) [cf.\ Fig.~\ref{figReplR2} (b,\ c)].  Inset: The asymptotic
  solution of $v_s$ for small values of $|\lambda-\lambda^*|$, where
  we set $\lambda^*=0$ for $d=1, 2$.}
\end{figure}

First, we consider the microorganism's motion with no fluctuations
present. In this case, the swimmer reaches a steady state at infinite
time, which is determined by a constant swimming speed $\dot{\bf
  r}_s(t\rightarrow\infty)={\bf v}_s$, where the index $s$ denotes the
steady-state configuration.  Under this condition, the drag force
induced by the solvent, $\gamma {\bf v}_s$, equals the driving force due
to the chemical, $-\lambda\nabla\rho$. Figuratively, the microorganism
surfs down its own chemorepellent, which it excreted at times
$t'<t-t_0$. Using Eq.~\eqref{eq:gradrho1}, the steady-state velocity
$v_s$ is determined by the self-consistent equation
\begin{eqnarray}\label{eq:vs}
\gamma v_s =
- \frac{\lambda v_s}{2D_c}\int_{t_0}^{\infty}{\mathrm d}t'
\frac{\exp\left[-v_s^{2} t'/(4 D_c)\right]}{(4\pi D_c t')^{d/2}}\nonumber\\
=\left(\frac{v_s}{4\pi^{1/2}D_c}\right)^d\frac{2\lambda}{v_s}
\,\Gamma\left(1-\frac{d}{2},\frac{v_s^2t_0}{4D_c}\right)\,,
\label{eq:gradrho2}
\end{eqnarray}
where $\Gamma(a,z)$ denotes the incomplete gamma function.
Eq.~\eqref{eq:vs} has non-zero solutions $v_s>0$ for any value of
$\lambda$ in one and two spatial dimensions, whereas there is a
dynamical phase transition at a lower critical value of
$\lambda^*= - 8\pi^{3/2}D_c^{5/2}t_0^{1/2}\gamma$ in three dimensions,
below which the microbe comes to a rest.  The latter is obtained
analytically by expanding Eq.~\eqref{eq:vs} up to second order in
$v_s$. For the parameters used in the simulations reported above,
$D_c=10$, $\gamma=10$, and $t_0=0.001$, the transition occurs at
$\lambda^*\approx - 4450$. The asymptotic solutions for small $|\lambda|$
in one and two dimensions or for small $|\lambda-\lambda^*|$ in three
dimensions, respectively, are given by
\begin{equation}\label{eq:vsasymp}
v_s(\lambda)\simeq\left\{\begin{array}{ll}
\left(2D_c\gamma\right)^{-1}|\lambda|\,,&d=1\\
2(D_c/t_0)^{1/2}\exp\left[-\frac{\gamma_E}{2}-\frac{4\pi D_c^2\gamma}{|\lambda|}\right]\,,&d=2\\
\left(4\pi^2D_c^2\gamma t_0\right)^{-1}\left(|\lambda-\lambda^*|\right)\,,&d=3\,,
\end{array}\right.
\end{equation}
where $\gamma_E \approx 2.7183$ is Euler's constant.  The steady-state velocity as a
function of $|\lambda|$ is shown in Fig.~\ref{fig:steadystate} for all
three spatial dimensions. For high coupling constants,
$\lambda\gtrsim10^4$, the steady-state velocities $v_s$ agree well
with the square root of the slope of the mean-square displacements
$v_s\approx {\partial{\sqrt{\langle({\bf r}_b(t)-{\bf r}_b(0))^2}}\rangle/\partial t}$, as obtained
from the simulations with noise in two and three dimensions
(see also Fig.~\ref{fig:steadystate}).

If noise is put on, again, the microbe is eventually disturbed in
its steady-state motion. In one dimension, the microbe needs to overcome a
barrier in order to change the direction of motion from left to right or vice
versa, which renders the steady state very stable. As we were not able to
observe long-time diffusive motion in one dimension for any coupling constant
but always found ballistic motion within the time window accessible in
computer simulations, we do not attempt to give a theoretical estimate of the
diffusivity. In two or three dimensions, the picture is very different: The
microorganism is only constrained in its motion parallel to its current
trajectory, whereas it is free to move perpendicular to the same. During such
transverse fluctuations the direction of the microbe's motion changes with
the gradient of the chemical field.

\begin{figure}
\includegraphics[width=8.0cm]{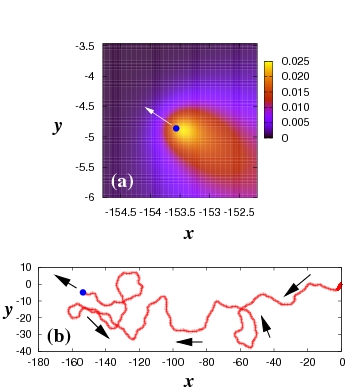}
\caption{\label{figRepChemDiffu} (color online).
(a) Snapshot of the instantaneous density profile $\rho(x,y)$ of
the chemorepellent released by the microorganism moving in $2-$dimensions,
obtained from simulation,
at time instant $t = 10.0$ (in units of ${\lambda_e}^{-1}$).
(b) The entire trajectory, shown as the red (thick) curve, of the
microorganism. The current position of the microorganism ${\bf r}_b (t)$
is indicated by the blue (black) dot in both the figures, and the direction
of motion is indicated by arrows along the trajectory. The corresponding
coupling strength being $|\lambda| = 10000$.}
\end{figure}

In the following, we give a theoretical estimate of the microbe's
long-time diffusivity int $d = 2, 3$ under the assumption of a fast
relaxation of the
chemical field, which is justified for $D_c\gg D$. In this case, the
change of orientation is determined by the local, time-independent
curvature $\kappa_0(\lambda)$ of the isodensity line (in 2D) or surface
(in 3D) at the steady-state position ${\bf r}_s(t)$; the isodensity
planes are defined by $\{{\bf r}^*(t)|\,\rho({\bf r}^*(t),t)=\rho({\bf
  r}_s(t),t)\}$. Locally, they have the form of a parabola (in 2D) or
of a paraboloid of revolution (in 3D) and they move with the microorganism
at the microorganism's velocity; this can be seen in
Fig.~\ref{figRepChemDiffu}, where a typical trajectory and the
according chemical field at time $t$ is plotted for a microorganism in two
dimensions.

Fluctuations transverse to the direction of motion lead to a mean
transverse displacement $\langle {\bf
  r}_{\perp}^2(t)\rangle=2(d-1)Dt$, which, in turn, leads to a change
in orientation of the velocity director ${\bf v}_b$ by a mean-square
angle
\begin{equation}
  \label{eq:meansquareangle}
 \langle\theta^2(t)\rangle=\kappa_0^2\langle {\bf
  r}_{\perp}^2(t)\rangle=2(d-1)D\kappa_0^2t \,.
\end{equation}
Exploiting that $\theta$ is Gaussian distributed, the average,
normalized, and on the initial orientation ${\bf v}_b(t=0)$ projected
velocity vector ${\bf v}_b(t)$ is given by
\begin{equation}\label{eq:averagecostheta}
  \frac{\langle{\bf v}_b(t)\cdot{\bf
      v}_b(0)\rangle}{v_s^2}=\langle\cos[\theta(t)]\rangle
  =\exp\left[-\frac{\langle\theta^2(t)\rangle}{2}\right]\,.
\end{equation}
Replacing time by arc length $s=v_st$, the problem under study can
therefore be mapped to the {\it worm-like chain} (WLC)
model~\cite{Kratky-Porod,doi86,Friedrich08} of a polymer of a total
arc length $L=v_s t$ and a persistence length, which is defined
implicitly by $\langle\cos[\theta(t)]\rangle=\exp[-v_st/l_p]$. Making
use of Eqs.~(\ref{eq:meansquareangle}) and (\ref{eq:averagecostheta})
the latter therefore reads $l_p= v_s/[(d-1)D\kappa_0^2]$. The mean
square end-to-end distance of the WLC for long chains $L\gg l_p$ is
well known~\cite{Kratky-Porod,doi86} to be given by
\begin{equation}
  \label{eq:WLC}
\langle{\bf r}_b^2(t)\rangle\simeq 2Ll_p=\frac{2v_s^2t}{(d-1)D\kappa_0^2} \,.
\end{equation}
The long-time diffusivity of the microbe is then obtained as the
time derivative of the mean square end-to-end distance,
\begin{equation}
  \label{eq:Dprime}
D_{l}(\lambda)=\frac{1}{2d}\frac{\partial \langle {\bf
  r}_{\perp}^2(t)\rangle}{\partial t}=\frac{v_s^2(\lambda)}{d(d-1)D\kappa_0^2(\lambda)}\,,
\end{equation}
where we point out the $\lambda$-dependence of the steady-state
velocity and the steady-state curvature. $D_{l}(\lambda)$ is plotted in
Fig.~\ref{figReplR2}~(d) and compared to the results of the
simulation.

It is ascertained from Fig.~\ref{figReplR2}~(d) that, in the limit of
large $\lambda\gtrsim10^5$, the theory overestimates the diffusivity
by a factor of $\sim3$, both in two and three dimensions, for the
parameters $D=10$ and $t_0=0.001$, almost independent of $\lambda$.
This discrepancy is attributed to the strong assumption of a constant,
non-fluctuating curvature $\kappa_0(\lambda)$ in
Eqs.~(\ref{eq:meansquareangle}) and (\ref{eq:Dprime}). Relaxing this
assumption to fluctuating curvature with first and second moments
$\langle\kappa\rangle=\kappa_0$ and
$\langle(\kappa-\kappa_0)^2\rangle=\Delta\kappa^2$, which is further
uncorrelated with the transverse displacement, i.e.,
$\langle\kappa(t){\bf r}_\perp(t')\rangle=0$, the diffusion constant
reduces to $D_{l}\rightarrow D_{l}/[1+(\Delta\kappa/\kappa_0)^2]$.
Therefore, although we do not attempt to give an estimate of
$\Delta\kappa$ here, the long-time diffusivity is expected to be
smaller than in the ``zeroth order'' theory, Eq.~(\ref{eq:Dprime}), in
agreement with the simulation results.



\section{\label{sec:level4}Conclusion}

In conclusion, we have explored the dynamics of autochemotaxis:
a model microorganism is ``smelling" its own secretion which is diffusing
away. The microorganism follows the gradient \cite{Endres} of its secreted
chemical. For the attractive case, the mean-square displacement of the
microorganism reveals a
transient dynamical arrest, most pronounced in low spatial dimensionality.
In the opposite case of chemorepulsion, there is a transient
ballistic behavior which crosses over to ultimate diffusion
\cite{note-evap,Wu}.
A simple theoretical analysis for large coupling strengths $\lambda$
reveals a scaling law for the long-time diffusion in the case of a
chemoattractant, by regarding small excursions of the microbe's position
about the bottom of an effective, time-dependent trapping potential.
In the case of chemorepulsion, the microbe's trajectory was mapped
to the contour of a worm-like chain, which gives a semiquantitative
agreement with our computer simulations.
The ranges of the coupling strength reported here are also easily obtainable
in real experimental situations ($\lambda \sim 10^{-1} - 10^4$), as
estimated earlier in the text. We note that though the mechanism of
temporal sensing \cite{kafri,Schnitzer,BergPurcell} of
chemoattractant or chemorepellent which we do not consider
here is found in many small, fast moving bacteria like {\em E. coli},
the direct gradient sensing mechanism is also present in microorganisms
like the amoeba {\em Dictyostelium discoideum}, the yeast cell
{\em Saccharomyces carevisiae}, lymphocytes, glial cells and
myxobacteria \cite{Luca,Endres,HaastertPostma}.

Chemotaxis in the gliding bacterium {\em M. xanthus} was demonstrated to be
in response to self-generated signaling chemicals \cite{kearns}.
However, the diffusivity
of the chemoattractant is significantly lower compared with the
bacterial motility in this peculiar case. A numerical investigation
\cite{tom} of a simple model case in one-dimension compared the interplay
between chemotaxis and
chemokinesis mechanism with a concentration dependent switching rate,
showing crossover from suppressed to enhanced diffusion in the parameter
space. This mean-field approach, studied in the limit of
vanishing chemical diffusivity and chemical degradation rate,
was based on the simplifying
assumption that the chemical coupling only affects the frequency of
direction reversal of the cell, keeping the cell speed unaltered.

Models of active colloids using surface reactions as a potential
mechanism for self-propulsion has been proposed
(see e.g. \cite{golestanian1,golestanian2}). In Ref.~\cite{golestanian1},
the model {\em molecular machine} is a spherical colloid
which reacts with the substrate at a specific site on its surface
and self-propel by asymmetric release of the reaction product.
The time evolution of the product particles is similar to that of
chemical molecules emitted by a microorganism. Though the exact way
of imparting a biased motion to the colloid by the emitted particles
is very different from chemotactic coupling, such systems
may also provide interesting situations to compare with our results.
In fact in Ref.~\cite{golestanian2}, active colloids interacting with a
self-generated cloud of solute were shown to have distinct propulsive
and anomalous super-diffusive regimes preceding a final long-time
effective diffusion. This is worth comparing with the ballistic to
a modified diffusion crossover in our study on negative autochemotaxis.

Future work should focus on generalizing the model to a
collection of model microorganisms \cite{geier2},
where hydrodynamic flow effects can play an important role
\cite{Howard,Hopkins,shapere} and steric repulsions can lead to
aggregation and clumps \cite{Holm}
as known from active particles \cite{TonerTuRamaswamy,Rik}.
It would further be interesting to study the case of a geometric confinement
of the particles \cite{Teeffelen,Hulme,Dietrich}  and the
chemoattractant/repellent in order to see effects
of a dimensional crossover. Finally, improving the theoretical approaches
towards full quantitative agreement will be an important task for future
research.

\begin{acknowledgments}
We thank M. Fuchs and R. Blaak for helpful discussions. This work was been supported by the SFB TR6
(DFG). 
\end{acknowledgments}

\bibliographystyle{apsrev}

\begin{thebibliography}{52}
\bibitem[{not()}]{note-evap}
\bibinfo{note}{The Green's function for a case with non-zero chemical
  extinction rate $(\Lambda)$ being known, it is not difficult to incorporate
  this into the equation of motion and simulate. The reason for considering
  $\Lambda = 0$ is not only to study a case which is experimentally realizable
  (as in Ref.\ \cite{Bonner}), but also to examine the dynamics of the microbe
  when the effect of the chemical is most strongly felt. Even in this limit we
  found a crossover to final diffusion. We therefore believe that the dynamics
  will not change qualitatively by setting a non-zero chemical extinction rate.
  Only the quantitative values of the crossover time scales, velocity and
  long-time diffusivity are expected to shift because of the reduced effect of
  the chemical.}
\end{thebibliography}

\end{document}